# Super-Planckian Radiative Heat Transfer between Macroscale Metallic Surfaces Due to Near-Field and Thin-Film Effects

Payam Sabbaghi, Linshuang Long, Xiaoyan Ying, Lee Lambert, Sydney Taylor, Christian Messner and Liping Wang*

*School for Engineering of Matter, Transport, and Energy*
*Arizona State University, Tempe, AZ, 85287, USA*

\* Corresponding author: [liping.wang@asu.edu](mailto:liping.wang@asu.edu)

**ABSTRACT**

In this study we demonstrate that the radiative heat transfer between metallic planar surfaces exceeds the blackbody limit by employing the near-field and thin-film effects over macroscale surfaces. Nanosized polystyrene particles were used to create a nanometer gap between aluminum thin films of different thicknesses from 80 nm to 13 nm coated on 5×5 mm$^2$ silicon chips, while the vacuum gap spacing is fitted from the near-field measurement with bare silicon samples. The near-field radiative heat flux between 13-nm-thick Al thin films at 215 nm gap distance is measured to be 6.4 times over the blackbody limit and 420 times to the far-field radiative heat transfer between metallic surfaces with a temperature difference of 65 K with receiver at room temperature. The experimental results are validated by theoretical calculation based on fluctuational electrodynamics, and the heat enhancement is explained by non-resonant electromagnetic coupling within the subwavelength vacuum gap and resonant coupling inside the nanometric Al thin film with *s* polarized waves. This work will facilitate the applications of near-field radiation in thermal power conversion, radiative refrigeration, and noncontact heat control where metallic materials are involved.

## I. INTRODUCTION

Photon tunneling in the near-field can enhance radiative heat transfer to overcome the blackbody limit governed by Planck's law when the vacuum gap between two radiating media is much less than the characteristic thermal wavelength [1, 2]. Potential applications of near-field radiation (NFR) include but are not limited to: near-field thermopower generation like thermophotovoltaics (TPV) [3-13], noncontact heat control like thermal rectification [14-17], and radiative refrigeration [18, 19], which have been widely discussed in recent years. It is well understood that the excitation of surface plasmon (SPP) or phonon polaritons (SPhP) [2, 20-23] with planar plasmonic or polar materials could significantly enhance near-field thermal radiation along with additional mechanisms like thin-film effect [6], hyperbolic modes [4, 24-29], magnetic polariton [30-35], and cavity resonance [36-38] recently identified with nanostructured materials.

Many reported near-field radiation experiments used polar materials [39-42] whose optical phonons in the infrared greatly enhanced the near-field radiative heat transfer near room temperature with coupled SPhPs to demonstrate super-Planckian radiative heat transfer with different experimental configurations at varied vacuum gap distances. AFM cantilever based sphere-surface techniques achieved gap distances down to 30 nm [43, 44], while the total radiative heat transfer is limited due to the relatively small surface area from the microspheres. Plate-plate setups with macroscale planar surfaces were another mainstream method, while the creation of nanometer gaps and parallelism across mesoscale lateral size is one of the major challenges [45-48]. Hu *et al*. [49] first used microscale polystyrene particles to separate two glass plates and measured a near-field enhancement of 50% over blackbody limit due to SPhP coupling. Ito *et al*. [50] formed a submicron gap using low-density pillars, and experimentally measured the radiative heat transfer between a pair of diced fused quartz substrates at a vacuum gap distance of 0.5 μm with a temperature differences up to 20 K. They also experimentally demonstrated a large amplitude modulation of radiation by the change of material properties across a 370 nm gap, which is maintained by the microfabricated spacers and applied pressure [51]. In addition, Ghashami *et al*. [52] developed a versatile near-field experimental setup based on a nanopositioning platform and observed more than 40 times enhancement of thermal radiation over blackbody radiation when two 5 × 5 mm$^2$ quartz plates were separated by a vacuum gap distance of 200 nm and with thermal gradients up to 156 K.

Semiconductors like silicon were also experimentally studied for near-field thermal radiation in particular with doping which could excite coupled SPPs in the infrared. For example, Shi *et al*. [53] tuned the nanoscale radiation between bulk silicon and a glass microsphere at a gap of ~60 nm by changing the carrier concentration of silicon. In addition, Lim *et al*. [54] measured the near-field thermal radiation between doped-Si parallel plates with a doping concentration of 8.33×10$^{19}$ cm$^{-3}$ by employing a MEMS-based platform. The near-field radiative heat transfer coefficient was found to be 2.91 times greater than the blackbody limit at a 400-nm vacuum gap. Watjen *et al*. [55] used SiO$_2$ posts between 1×1 cm$^2$ lightly doped silicon samples with a doping concentration of 2×10$^{18}$ cm$^{-3}$. The largest radiative heat flux was found to be 11 times higher than the blackbody limit at a vacuum gap distance of 200 nm, which was mainly attributed to the excitation of coupled SPPs. Bernardi *et al*. [56] demonstrated a radiative heat transfer enhancement of 8.1 times relative to the blackbody limit between two 5×5 mm$^2$ intrinsic silicon planar surfaces at a vacuum gap distance of 200 nm, which is due to the additional contribution of frustrated modes in the near-field. More recently, Ying *et al*. [57] created vacuum gaps by SU-8 polymer posts, and experimentally found the near-field thermal radiation enhancement over the blackbody limit by 11 times between highly doped silicon chips at 190 nm gap spacing. In addition, DeSutter *et al.* [58]

measured a maximum heat transfer enhancement of approximately 28.5 over the blackbody limit with devices made of millimeter-sized doped Si surfaces separated by vacuum gap spacings down to approximately 110 nm.

On the other hand, metals are much less studied for near-field thermal radiation, as it is known that their plasma frequencies typically exist in the ultra-violet to the visible wavelength ranges [59] and therefore coupled SPP cannot occur in the infrared for significant near-field enhancement around room temperature. In addition, metals are normally known as bad thermal emitters in the far-field due to their very low emissivity, typically just a few percent. However, in numerous near-field heat transfer systems, metallic surfaces are involved or desired as high-temperature thermal emitters or electrodes in particular for thermophotovoltaic devices. Shen *et al*. [60] measured the enhanced thermal radiation due to the tunneling of non-resonant evanescent waves between a 50 μm diameter glass microsphere nominally coated with a 100 nm-thick gold film and a substrate coated with a thick gold film at a vacuum gap distance of 30 nm. Kralik *et al*. [61] observed a one hundred fold enhancement of the blackbody radiation when the plane-parallel tungsten layers were separated by a vacuum gap distance of 1 μm, at cryogenic temperatures (i.e., 5 K to 40 K). Lim *et al*. [62] measured the radiative heat flux between metallo-dielectric multilayers at a vacuum gap distance of 160 nm, achieving a net radiative heat flux more than 100 times larger than the calculated far-field value and about seven times larger than the blackbody limit due to coupled SPPs at metal-dielectric interfaces. Song *et al*. [40] also reported the near-field enhancement between silicon microdevices at a size of 48×48 μm$^2$ coated with 100 nm-thick gold layer. The near-field thermal conductance at a gap distance of 60 nm was found to be ~10 times larger than the blackbody limit.

In this study, we experimentally and theoretically demonstrate super-Planckian radiative heat transfer between ultra-thin aluminum films due to near-field and thin-film effects. After a careful cleaning process using isopropanol alcohol, deionized water, and oxygen plasma, Al thin-films of different thicknesses were coated on 5×5 mm$^2$ lightly doped Si chips (Ted Pella Inc., 1-30 ohm-cm resistivity, 500±30 μm thick) via an electron beam evaporation method with the deposition rate of 0.5 Å/s. The thicknesses of the deposited aluminum thin-films were characterized to be 13±2 nm, 24±3 nm, 40±3 nm, and 79±3 nm by an atomic force microscope along with surface roughness of 0.4 nm as shown in Fig. S1 (supplemental material). The bow of the Si chip was measured to be 75±19 nm with a profilometer. A polystyrene nanoparticle suspension in deionized water (Sigma Aldrich, 69057-5ML-F, with a nominal diameter of 200 nm) was diluted to the desired concentration of 7.8×10$^6$ particles/mL. Then, 0.02 mL of the diluted suspension was carefully distributed using a syringe on the receiver samples, which were then dried on a hotplate (Section B in the supplemental material). The receiver samples with nanoparticles were carefully inspected under an optical microscope to ensure that were no large particle aggregations (Fig. S2 in the supplemental material) before being loaded onto the experimental setup for near-field radiation measurements.

## II. EXPERIMENTAL NEAR-FIELD SETUP AND HEAT TRANSFER MODEL

Figure 1(a) illustrates the experimental platform developed to measure the near-field radiative heat flux, where an Al thin-film emitter with the same thickness as the receiver is separated from the receiver sample by the polystyrene particles, under the total applied force of 30 mN which creates a vacuum gap around 200 nm. A thermoelectric heater is used to vary the emitter sample temperature, while a thermoelectric cooler maintains the receiver sample at a constant temperature of 23°C to create temperature differences, $\Delta T$, from 25 K to 65 K by adjusting power inputs independently with different DC power supplies. Copper plates with 3.25 mm thickness and the conduction thermal resistance, $R_{copper} = 0.325$ K/W, are used to spread the heat uniformly underneath (above) the receiver (emitter) samples. The temperatures of the emitter ($T_1$) and receiver ($T_2$) are measured by thermistors embedded inside the center of the Cu plates with an accuracy of ±0.1°C. Thermal grease (Arctic Silver Ceramique 2) is used at every interface to minimize contact resistance $R_{grease}$, which varies from 0.129 K/W to 0.328 K/W between measurements. A 1.1-mm-thick glass microscope slide with a thermal conductivity of 1 W m$^{-1}$ K$^{-1}$ and conduction thermal resistance, $R_{glass} = 44$ K/W, is placed below the receiver sample, in order to find the amount of heat transfer by measuring the copper plate temperatures $T_2$ and $T_3$ across the slide. Note that the copper plates, samples, and glass slide have the same area of $A = 5 \times 5$ mm$^2$. According to the 1D thermal resistance network in Fig. 1(b), the rate of heat transfer $Q_{glass}$ across the glass slide can be calculated as

$$Q_{glass} = \frac{T_2 - T_3}{2R_{grease} + R_{copper} + R_{glass}} \quad (1)$$

Under steady-state conditions and by neglecting side losses, the energy balance yields $Q_{total} = Q_{glass} = Q_{rad,exp} + Q_{cond,PS}$, where $Q_{cond,PS}$ is the heat conduction via the polystyrene (PS) particles. Therefore, the rate of near-field radiative heat transfer $Q_{rad,exp}$ from the experiment can be determined by

$$Q_{rad,exp} = Q_{total} - Q_{cond,PS} \quad (2)$$

With the given nanoparticle concentration, $Q_{cond,PS}$ is less than 8% of the theoretical near-field radiative heat transfer between lightly doped Si at a vacuum gap distance of 200 nm (Fig. S6a in the supplemental material).

The experimental setup was placed inside a 12-inch diameter vacuum chamber (Fig. S3 in the supplemental material) where the measurement was conducted in a high vacuum environment provided by a turbo pump with a pressure less than 0.1 Pa. At this pressure, the conduction heat transfer with air molecules is less than 0.2% of the near-field radiative heat transfer between lightly doped Si at 200-nm vacuum gap from calculation (Fig. S6b in the supplemental material). The temperature difference between the emitter and receiver samples is obtained by tuning both the TE heater and cooler, while the $T_1$, $T_2$, and $T_3$ readings were recorded for each measurement at steady state. The measurement is considered to be at steady-state when the variation of all temperatures was less than 0.1°C over 5 minutes. Three independent measurements were conducted for each Al thin-film of the same thickness, with a new pair of samples being used for each subsequent measurement. To ensure the consistency of the experimental procedures, the same suspension of PS nanoparticles was distributed over the Al thin-film samples with the same thickness that were fabricated from the same batch during the same particle distribution process along with one pair of bare Si chips.

One major challenge in the plate-plate near-field radiation measurements with mm-sized samples is how to experimentally determine the gap distance accurately. The optical interferometry method, which requires double-side polished semitransparent samples, has been used to determine the gap distance by fitting the interference fringe measured between lightly doped Si before near-field measurement [55] and between quartz samples during the near-field measurement [50, 52]. However, it cannot be applied to metallic thin-films due to opacity or very small transmission due to strong light attenuation. Also, the nominal diameter of the nanoparticles cannot be discounted as a precise vacuum gap distance. Here, we conducted additional near-field radiation measurements with 4 pairs of bare Si chips, each coated with the same nanoparticle suspension at the same time with Al thin-film sample of different thickness. The optical constants used for Si in the calculation were obtained from Palik [63] (Section D in the supplemental material). The gap distance was obtained by fitting the measured near-field radiative heat flux between Si plates $q_{\text{rad,exp}} = Q_{\text{rad,exp}}/A$ with theoretical values $q_{\text{rad,theo}}$ calculated by fluctuational electrodynamics (Section E in the supplemental material). To ensure the accuracy of the optical constants, spectral reflectance and transmittance of a double-side polished Si wafer with the same thickness and resistivity range were measured with an infrared spectrometer. The wavelength-dependent optical constants of lightly doped Si were extracted with the Ray-tracing method [1], which showed excellent agreement with Palik data [63] in Fig. S4 (supplemental material). In addition, extended measurement uncertainty from four independent experiments and accuracy from error propagation, was considered for the experimental near-field radiative heat flux from bare Si chips. A least-square method (Section G in the supplemental material) was used to fit the vacuum gap to be $d = 215^{+55}_{-50}$ nm according to the upper bound, average, and lower bound of the measured $q_{\text{rad,exp}}$ for Si at different $\Delta T$ values in Fig. S7 (supplemental material).

## III. RESULTS AND DISCUSSION

The comparison between the measured near-field radiative heat flux and the theoretical calculations at the fitted vacuum gap of $d$ = 215 nm is shown in Figs. 2(a-d) for Al thin-films with different thicknesses: (a) 13±2 nm, (b) 24±3 nm, (c) 40±3 nm, and (d) 79±3 nm. The shaded area displays the calculated near-field radiative flux considering the uncertainty of the vacuum gap distance $d = 215^{+55}_{-50}$ nm obtained from four near-field radiation measurements between bare Si chips, whereas solid black lines denote the near-field radiative heat fluxes at the average vacuum gap distance of $d$ = 215 nm. The measured near-field radiative heat fluxes $q_{\text{NFR,exp}}$, symbolized by the red markers with error bars as combined uncertainties $U_c$, overlap with theoretical predictions considering upper and lower bounds due to the uncertainty of the fitted vacuum gap, indicating good agreement. There is slight difference in particular for 13-nm Al thin film, which is mainly because of optical constants of Al used for modelling and experimental uncertainties of vacuum gap and heat flux measurements.

It can be observed that the near-field radiative heat flux increases as the thickness of aluminum thin-film decreases. It should be noted that the effect of native oxidation of Al with 3-nm $Al_2O_3$ on the total near-field radiative heat flux is negligible based on the calculations not shown here. In the insets of Figs. 2(a-d), the total radiative heat flux of the Al thin films at a vacuum gap distance of 215 nm is compared with the blackbody limit and far-field radiation of bulk aluminum surfaces. The theoretical results indicate that the total radiative heat flux between the 13-nm Al thin-film samples separated by a vacuum gap of 215 nm is approximately 5224 W/m$^2$ for the temperature difference $\Delta T = 65$ K, indicating a theoretical improvement about 10 times over the blackbody limit and 650 times compared to the far-field radiation of bulk aluminum sample. The near-field radiative heat flux for thicker Al samples was calculated to be 3943 W/m$^2$, 2846 W/m$^2$ and 1901 W/m$^2$ for 24, 40 and 79-nm-thick Al films, respectively, under the same temperature difference of 65 K. As shown in Fig. 2, the near-field radiative heat flux decreases monotonically with increasing aluminum thickness.

Near-field radiative heat flux enhancement factors over the blackbody limit and over far-field radiation of bulk aluminum plates as a function of aluminum thickness and for a temperature difference of 65 K are plotted in Figs. 3(a) and (b), respectively. At a vacuum gap distance of 215 nm, the near-field radiation heat flux was experimentally measured with an average enhancement of 6.4 times over blackbody limit and ~420 times over far-field radiation with bulk Al, which is roughly consistent for all four Al thin-films of different thicknesses of 13 nm, 24 nm, 40 nm and 79 nm. On the other hand, calculations predict that the largest near-field radiative enhancement at $d = 215$ nm occurs with the 5-nm-thick Al thin-film attaining 11 times over the blackbody limit and 720 times over far-field radiation of bulk aluminum plates, respectively. Due to the uncertainty with the near-field heat flux measurement as well as with the theoretical prediction from the fitted vacuum gap $d = 215^{+55}_{-50}$ nm, the experimental data and calculation are considered to be within reasonable agreement as shown in Fig. 3. In addition, the insets display the near-field enhancement factor over the blackbody limit and over far-field radiation of bulk aluminum calculated at smaller vacuum gap distances of 50 nm and 100 nm, respectively. The results suggest that the near-field enhancement over the blackbody limit could potentially reach 42 times with 15-nm-thick Al at a vacuum gap of 100 nm, or even 123 times with 20-nm-thick Al at a vacuum gap of 50 nm. When compared to the far-field radiation with bulk Al, the near-field and thin-film effects could enhance the radiation heat flux up to 2750 times with 15-nm-thick Al at $d = 100$ nm or 8060 times with 20-nm-thick Al at $d = 50$ nm.

In order to understand the mechanism responsible for higher total heat flux with the Al thin film than the bulk, the energy transmission coefficient $\xi$ for $s$ polarization only at $d = 215$ nm gap distance was plotted in Figs. 4(a) and (b), respectively for 13-nm Al and the bulk as a function of frequency $\omega$ and normalized wavevector $\beta c_0/\omega$. A broadband enhancement with close-to-unity transmission coefficient at frequencies less than $1\times10^{13}$ rad/s and normalized wavevectors up to 20 or so is clearly seen for 13-nm thin film only but not for the bulk, which indicates the thin-film effect. The reflection coefficient at the vacuum interface (Eq. S3 in the supplemental material) for the 13-nm Al thin film on Si substrate has a Fabry-Perot-like denominator $DE = 1 + r^s_{01} r^s_{12} e^{i2\gamma_1 t}$, where indices 0, 1, and 2 are for vacuum, Al thin film, and Si substrate, respectively. At the low frequency range, the real part of permittivity of aluminum is about −10000 from the Drude model in Fig. S5(a) (supplemental material), leading to $\gamma_1 = \sqrt{\varepsilon_{Al}k_0^2 - \beta^2} \approx i100k_0$ (when $\beta < 20k_0$) and $\gamma_1 \gg \gamma_0$ or $\gamma_2$, such that $r_{01}^s = \frac{(\gamma_0 - \gamma_1)}{(\gamma_0 + \gamma_1)} \to -1$ and $r_{12}^s \to 1$. Besides, when the Al thin film thickness $t$ is around 15 nm or less, the exponent term $e^{i2\gamma_1 t}$ approaches 1, and therefore the denominator $DE$ goes to zero, which indicates resonant coupling within the Al thin film. This is verified by plotting the $DE$ expression for the 13-nm Al thin film in Fig. 4(c), where it is almost zero at these low frequency $\omega$ ranges and small wavevector $\beta$ values. On the other hand, the denominator of the reflection coefficient for the bulk Al (i.e., $r_{01}^s$ at the single vacuum-Al interface), is also presented in Fig. 4(d) for comparison, where the large values do not suggest any resonance at all for bulk Al. Note that the near-field spectral radiative heat fluxes of $s$ and $p$ polarizations at 215 nm gap were calculated in Fig. S8, along with the effect of vacuum gap distance on the total near-field radiative heat flux in Fig. S9 (supplemental material), to further illustrate the heat transfer enhancement with ultra-thin aluminum films over bulk Al plates.

## IV. CONCLUSION

In summary, this work demonstrated the near-field and thin-film effects on enhanced radiative heat transfer between metallic surfaces with thin aluminum films coated on silicon chips at a vacuum gap distance of 215 nm. The near-field radiative heat flux between 13-nm-thick Al films was experimentally measured at a vacuum gap distance of 215 nm with an enhancement of 6.4 times over the blackbody limit and 420 times over that of the far-field radiative heat transfer between bulk aluminum at a temperature difference of 65 K with receiver at room temperature. Theoretical predictions from fluctuational electrodynamics validated the experimental results with the reasonable agreement, and clarified the large enhancement was because of augmented contributions from *s* polarization with non-resonant coupling within the subwavelength vacuum gap (i.e., near-field effect) and resonant coupling within the nanometric Al thin film (i.e., thin-film effect). The advances of experimental measurements and fundamental understanding of near-field radiation between metallic surfaces here could greatly facilitate the applications of various near-field thermophotonic devices for noncontact thermal energy conversion and heat control.

**Supplementary Material:** Supporting information includes the details on sample fabrication and characterization, nanoparticle preparation, optical properties, theoretical models, conduction heat transfer calculation, fitting of vacuum gap, spectral radiative heat flux, and vacuum gap effect along with Figs. S1-S9.

**Data Availability:** The data that support the findings of this study are available from the corresponding author upon reasonable request.

**Acknowledgements:** This work was mainly supported by Air Force Office of Scientific Research with Grant No. FA9550-17-1-0080 (P.S., X.Y., and L.W.) and National Science Foundation under Grant No. CBET-1454698 (L.L and L.W.). Access to the ASU NanoFab for sample fabrication was supported in part by NSF contract ECCS-1542160. Support from ASU Fulton Undergraduate Research Initiative was also appreciated.

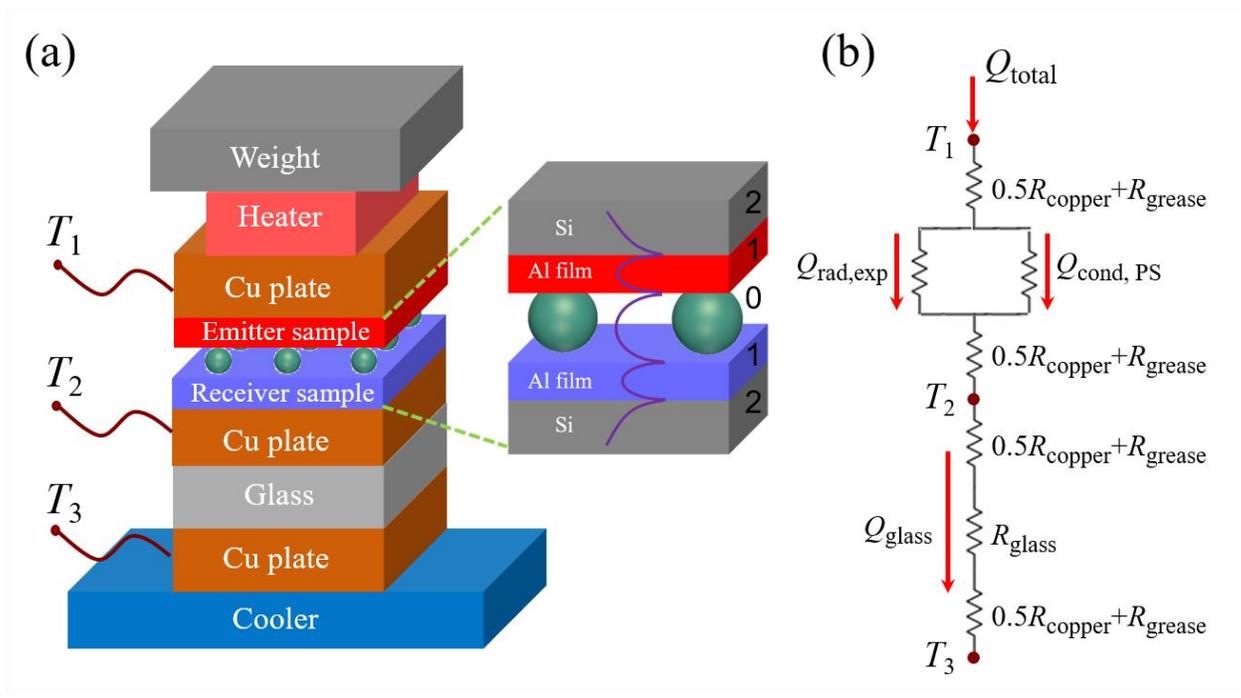

FIG. 1. (a) Schematic of the experimental setup with the illustration of coupled electromagnetic waves inside the vacuum gap and within the Al thin film. Thermistors were inserted inside the Cu plates to measure the temperatures. The emitter sample is maintained at temperature of $T_1$ and the receiver is at $T_2$ separated by a vacuum gap distance $d$. (b) Equivalent thermal circuit model to illustrate the 1D heat transfer.

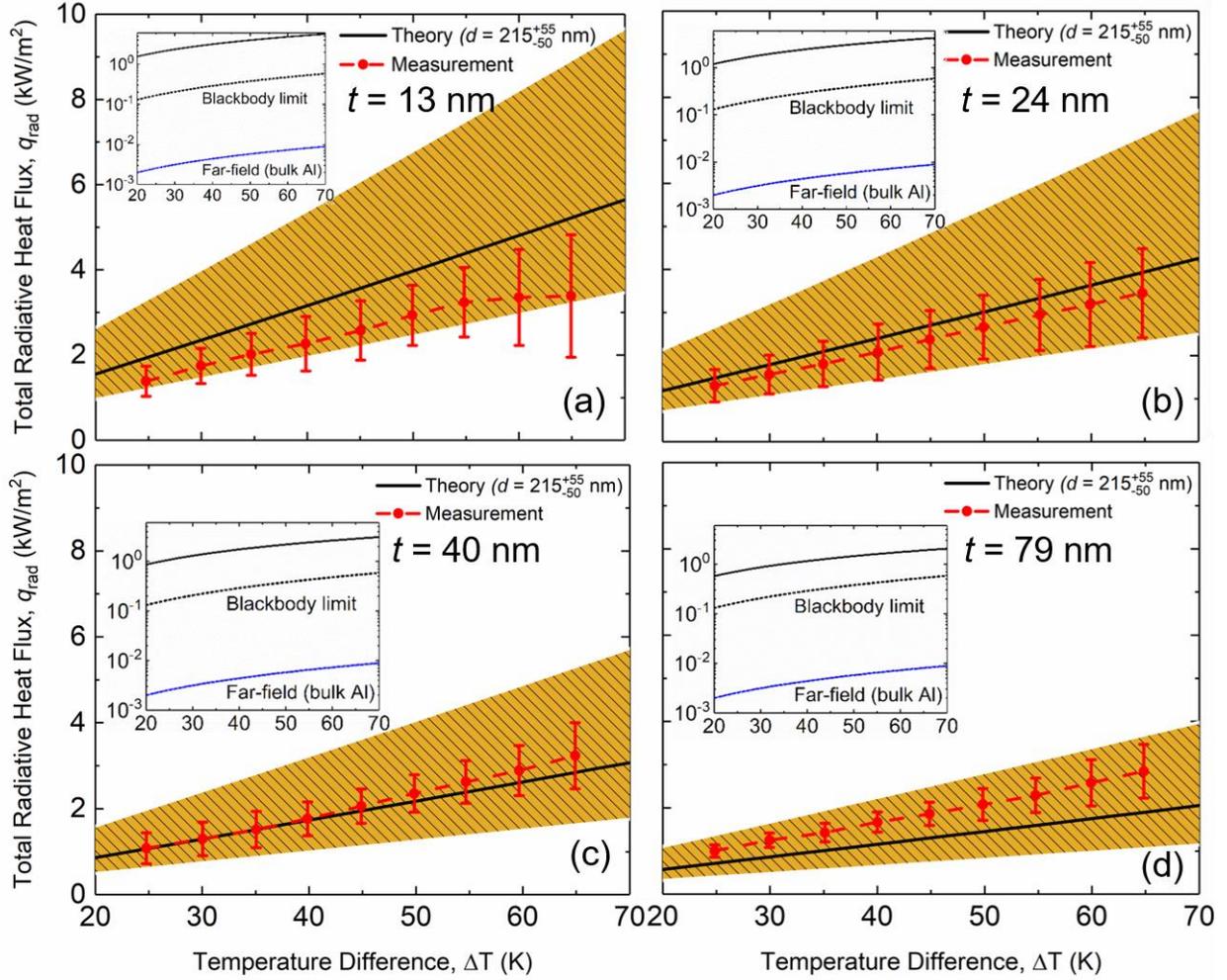

FIG. 2. Measured near-field radiative heat flux (markers with error bars) between Al thin films of several thicknesses at different $\Delta T$ values along with the theoretical prediction (shaded) at a vacuum gap distance $d = 215^{+55}_{-50}$ nm fitted from the near-field measurement with bare Si chips. The thicknesses of the aluminum are: (a) 13±2 nm, (b) 24±3 nm, (c) 40±3 nm, and (d) 79±3 nm. The inset shows the total radiative heat flux at a vacuum gap distance of 215 nm compared with the blackbody limit and far-field radiative heat transfer of bulk aluminum plates.

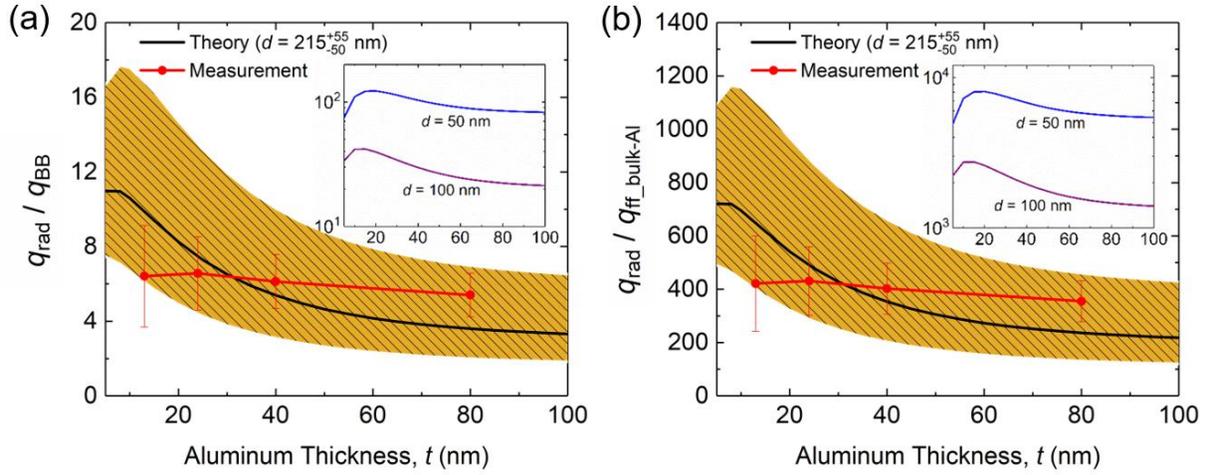

FIG. 3. Measured near-field heat flux enhancement (markers with error bars) between Al thin films at different thicknesses compared to (a) blackbody limit and (b) far-field radiative heat transfer between bulk aluminum plates. The shaded region indicates the vacuum gap uncertainty $d = 215^{+55}_{-50}$ nm while the average vacuum gap of 215 nm is shown by the solid black line. The insets inside (a) and (b) show the total radiative heat flux enhancement over blackbody and far-field limits at vacuum gap distances of 50 nm and 100 nm, respectively. Note that the temperature of the receiver is kept constant at 23°C and the temperature difference is 65 K.

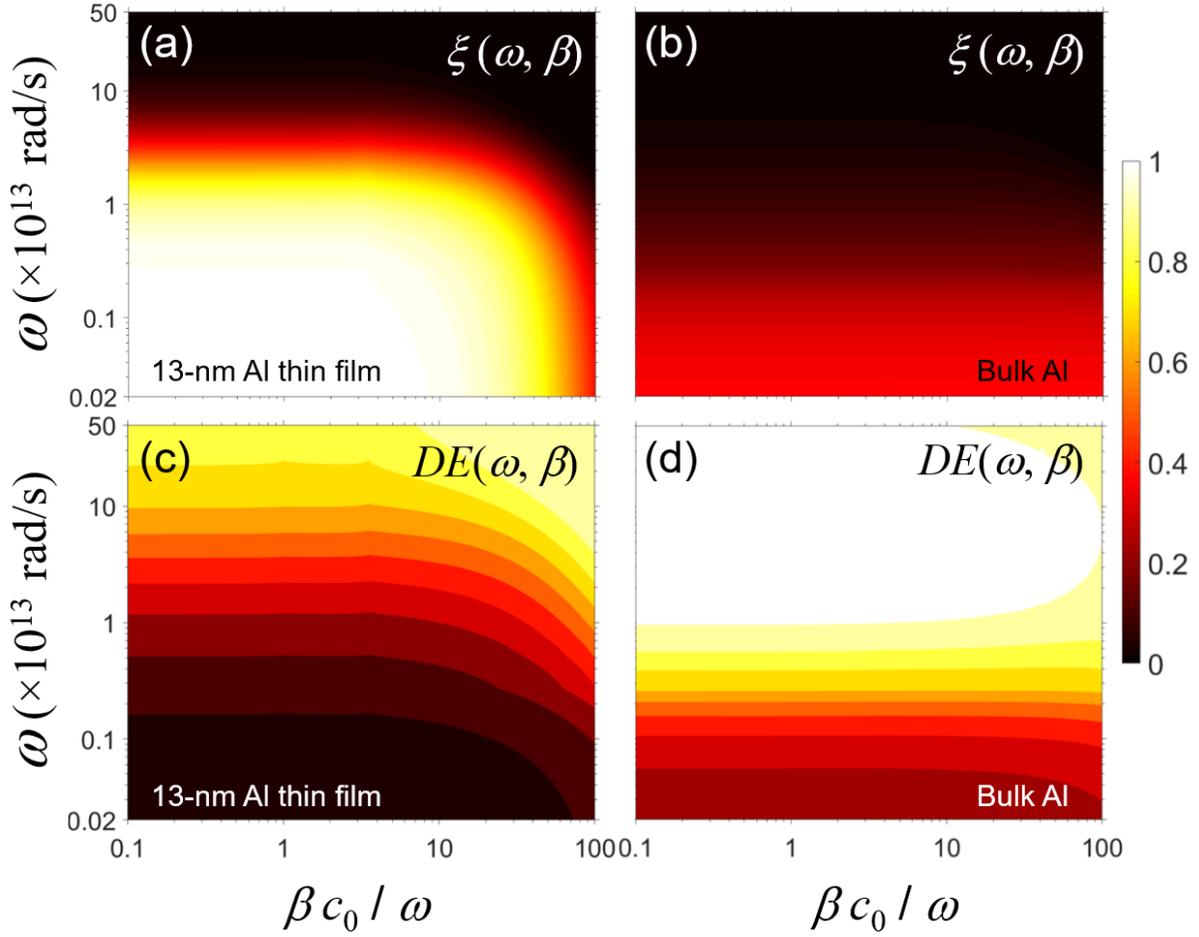

FIG. 4. Contour plot of energy transmission coefficient ($\xi$) for $s$ polarization only at $d = 215$ nm gap distance (a) between lightly doped silicon samples coated by 13 nm aluminum, and (b) between bulk aluminum samples. Denominator of the reflection coefficient are also shown for (c) the 13-nm Al thin film, and (d) bulk Al. Note that the parallel wavevector component is normalized to the frequency.

18